\documentclass[twocolumn, reprint,
showpacs,
preprintnumbers,
amsmath,
amssymb,
amsfonts,
final,
a4paper,
aps,
floatfix,
citeautoscript,
footinbib,
prl,
superscriptaddress]{revtex4}
\usepackage[linkcolor = red, citecolor = blue, urlcolor = blue]{hyperref}
\usepackage{times}
\usepackage{graphicx}
\usepackage[latin1]{inputenc}
\usepackage{color}
\usepackage{multirow}
\usepackage[textwidth=175mm,textheight=265mm]{geometry}
\newcommand{\bq}{\mathbf{q}}
\newcommand{\bk}{\mathbf{k}}
\newcommand{\bl}{\mathbf{l}}

\begin{document}
\title{Phenomenological Characterization of Semiholographic Non-Fermi Liquids}
\date{\today}
\author{Ayan \surname{Mukhopadhyay}}
\affiliation{Centre de Physique Th\'{e}orique, \'{E}cole Polytechnique, CNRS, 91128 Palaiseau, France}
\email{ayan.mukhopadhyay@cpht.polytechnique.fr}
\affiliation{Institut de Physique Th\'{e}orique, CEA Saclay, CNRS URA 2306, F-91191 Gif-sur-Yvette, France}
\author{Giuseppe \surname{Policastro}}
\affiliation{Laboratoire de Physique Th\'eorique, Ecole Normale Sup\'erieure, 24 rue Lhomond, 75231 Paris, France} 
\email{policast@lpt.ens.fr}
\pacs{11.25.Tq, 71.10.Hf, 71.27.+a}
\begin{abstract}
We analyze some phenomenological implications of the most general semiholographic models for non-Fermi liquids that have emerged with inputs from the holographic correspondence. We find generalizations of Landau-Silin equations with few parameters governing thermodynamics, low-energy response and collective excitations. We show that even when there is a Fermi surface with well-defined quasiparticle excitations, the collective excitations can behave very differently from Landau's theory. 
\end{abstract}
\maketitle


{\it Introduction.} In recent years, a large number of experiments conducted on different metals, for instance, underdoped  cuprate superconductors and heavy fermion materials, have observed behaviour in terms of transport and thermodynamic properties \cite{stewart} that could not be explained by conventional Landau theory of Fermi liquids. These ``non-Fermi" liquids present a challenge for theorists, and so far, there is no accepted theory, although several models have been proposed (see e.g. Ref. \cite{varma}).  For instance, if one couples an ordinary Fermi liquid to a  
gapless system, which could be a gauge theory or a system near a critical point, the long-range critical fluctuations turn the ground state into a non-Fermi liquid \cite{NFL}, although the system cannot be studied in any controlled approximation. 

In Fermi liquids, the gapless excitations around the Fermi surface have the same quantum numbers as the free fermion. The effective theory describing them can be understood as a fixed point of a Wilsonian RG. In non-Fermi liquids there is a Fermi surface but the gapless excitations are not adiabatically connected to the free fermion states. The low-energy dynamics cannot be readily understood as a fixed point of Wilsonian RG, since the nontrivial scaling exponents of the Fermi surface are naturally explained only by a nonlocal effective theory (see a recent discussion in \cite{Fitzpatrick:2013mja}).

In such a situation, one can gain insights from solvable models. Holographic duality is the only nonperturbative tool available for strongly interacting fermions at finite density, though the class of field theories for which this works is indeed unrealistic.
The duality maps states of a strongly coupled field theory in $d$ dimensions to solutions of a weakly coupled gravity or string theory in $d+1$ dimensions  \cite{adscft}.  In particular, a finite density state at zero temperature maps to an extremal black hole. The duality makes generic predictions on the features of systems at a quantum critical point (e.g., quantum dissipation or charge fractionalization \cite{Hartnoll:2007ih, reviews})   that can be matched to those observed in real materials or in field-theoretic models  \cite{sachdev}. In particular, the phenomenological form of the quasiparticle spectral function near the Fermi surface, with nontrivial scaling exponents, has been reproduced using the fermionic response of a dual extremal black hole  
 \cite{Liu:2009dm, Iqbal:2011ae}. This discovery was certainly quite exciting: one can obtain continuous families of solvable models for non-Fermi liquids whose properties can be easily investigated. At special points in the parameter space, one obtains features resembling those of a marginal Fermi liquid. 
 
In these holographic models, the low-energy form of the spectral function,  and thus the scaling exponents,  are completely fixed by the near-horizon geometry of the black hole. This near-horizon geometry has a scaling symmetry and represents an emergent infrared CFT, which could belong to an universality class that includes realistic materials, even though the underlying microscopic holographic theory was unrealistic. 

Building on previous suggestions  \cite{Hartnoll:2009ns}  Faulkner and Polchinski \cite{Faulkner:2010tq} have proposed a minimalistic framework to describe all possible such emergent infrared CFTs and the related non-Fermi liquid state in a setup that is semiholographic.  Only the infrared degrees of freedom are holographically represented by the near-horizon region of the classical geometry,  while the ultraviolet degrees of freedom are represented by the weakly coupled fermionic field $\chi$ living at the boundary of the space and their dynamics is described perturbatively. Furthermore $\chi$ couples linearly to the composite operator $\psi$ with the same quantum numbers in the holographic CFT whose dynamics is captured by $S_{\rm{hol}}$ which is the on-shell gravitational action of the near-horizon geometry. 
The action describing the model is: 
 \begin{eqnarray}\label{modeltrunc}
 S &=&  \int dt \sum_k \Big[\chi^{\dagger}_{\mathbf{k}}(i\partial_t - \epsilon_{\mathbf{k}}+\mu)
\chi_{\mathbf{k}} 
+  ( g_{\mathbf{k}} \chi^{\dagger}_{\mathbf{k}}
\psi_{\mathbf{k}} + c.c.  \Big)\Big] \nonumber \\
&+&  S_{\textrm{hol}} \,. 
\end{eqnarray}

This model  captures  the infrared physics of the fully holographic constructions and at the same time allows for many extensions; for instance, the interior geometry can be taken to be $AdS_4$,  $AdS_2 \times {\mathbb R}^2$, a geometry with Lifshitz scaling and/or violating  hyperscaling, depending on the different types of infrared CFTs. In each case, the model describes an IR fixed point that is expected to encode the universal properties of a large class of interacting fermion systems. One could also introduce a lattice through a $\mathbf{k}$ dependence in $g_\mathbf{k}$ or take into account other realistic complications.
We notice also that the mixing of the ``electron" $\chi$ with the fermion operator of the CFT is reminiscent of the phenomenon of \emph{fractionalization} that plays a role in the description of strange metals.  

In the present Letter, we want to extend this phenomenological framework beyond the simple deconstruction of the holographic setup. For this purpose we need to consider higher-order interaction terms. They will be generically present if the model is considered as an effective theory derived by integrating out high-energy modes. Moreover they are needed to account for collective modes of fermions, such as zero sound, and for the electrons to give nontrivial contributions to the thermodynamic and transport properties, even though only at subleading order for large $N$. 
 
An important caveat in the models we consider is that the degrees of freedom of the boundary fermions are parametrically fewer than those of the soft modes that admit a 
 holographic description. Without this assumption, one would have to take into account the backreaction of the fermions  \cite{Hartnoll:2009ns, Hartnoll:2010gu, Cubrovic:2010bf, Sachdev:2011ze, Allais:2012ye, Medvedyeva:2013rpa}, which would completely change the IR fixed point. 
 
We will generalize the standard Landau-Silin theory of normal FL in our models allowing us to obtain phenomenological predictions for collective modes and transport properties in terms of analogues of Landau parameters. The latter are defined microscopically in terms of quasiparticle scattering at the Fermi surface. 
We will then explicitly solve for the zero sound mode dispersion relation and the plasma frequency. We end with a discussion and point out future directions of investigation. 

{\it The model.} 
We consider the following action that describes the most general semiholographic model, including the possibly relevant or marginal perturbations that can be constructed from the boundary fermions and the CFT operators (see the Supplementary Material for a discussion of the RG in Fermi liquids and in this model)  : 
\begin{widetext}
 \begin{eqnarray}\label{model}
&&  S = \int dt \Bigg[ \sum_k \Big(\chi^{\dagger}_{\mathbf{k}}(i\partial_t - \epsilon_{\mathbf{k}}+\mu)
\chi_{\mathbf{k}} 
+\frac{1}{2}   \sum_{k,k_1,q} \chi^\dagger_{\mathbf{k}}
\chi_{\mathbf{k}-\mathbf{q}}V(\mathbf{q})
\chi^\dagger_{\mathbf{k}_1}
\chi_{\mathbf{k}_1-\mathbf{q}}
+  \sum_{k,k_1,q}\lambda_{\mathbf{k}, \mathbf{k}_1, \mathbf{q}}\chi^\dagger_{\mathbf{k}}
\chi_{\mathbf{k}-\mathbf{q}}
\chi^\dagger_{\mathbf{k}_1}
\chi_{\mathbf{k}_1-\mathbf{q}}  
\\
&& + N  \sum_{k} ( g_{\mathbf{k}} \chi^{\dagger}_{\mathbf{k}}
\psi_{\mathbf{k}} + c.c.  \Big)
+N   \sum_{k,k',q}  \eta_{\mathbf{k},\mathbf{k}'}\chi^{\dagger}_{\mathbf{k}}
\chi_{\mathbf{k}'}
\phi_{\mathbf{k}-\mathbf{k}'}
+ N  \sum_{k,k_1,k_2}
\Big(\tilde{g}_{\mathbf{k}, \mathbf{k}_1, \mathbf{k}_2}\chi^{\dagger}_{\mathbf{k}} 
\chi_{\mathbf{k}_1}
\chi^{\dagger}_{\mathbf{k}_2}
\psi_{\mathbf{k}-\mathbf{k}_1+\mathbf{k}_2}
+c.c. \Big) \Bigg]
+ N^2 S_{\textrm{bulk}} \,. \nonumber
\end{eqnarray}
\end{widetext}
In comparison to Eq. (\ref{modeltrunc}) we consider both bosonic and fermionic CFT operators ($\phi$ and $\psi$ respectively) that can couple to $\chi$ by cubic or quartic interactions. We allow also for self-interactions of $\chi$ and for convenience we separate, in the quartic self-coupling,  the long-range interactions (e.g. Coulomb) in $V(q)$ from the short-ranged ones in $\lambda_{\bk,\bk_1,\bq}$ that are assumed to be a nonsingular functions of the momenta.  We have also included a parameter $N$ that allows us to have a parametric control of the diagrammatic expansion. We stress that it is not related to the central charge of the bulk CFT, which is typically a $U(N_c)$ gauge theory with $c \sim {\cal O}(N_c^2)$. 

We adopt an $N$ counting that realizes the following features: (i) the corrections to the $\chi$ propagator due to the bilinear coupling with $\psi$ are ${\cal O}(1)$, so that at leading order 
we reproduce the non-Fermi liquid form of the propagator of Ref. \cite{Faulkner:2010tq}, (ii) loops of the boundary fermion are not suppressed, as indeed we are interested in their full dynamics, (iii) loops in the bulk are suppressed, and (iv) a connected $n-$point bulk tree diagram attached to boundary $\chi$-fermion lines is of order $N^{2-n}$, thus leaving only bulk propagators unsuppressed while vertices in the bulk are suppressed. These make the model tractable. 

After resumming the  geometric series for the propagator, we find that  only the $\chi-\chi$ propagator has a Fermi-surface singularity, with a form depending on the IR geometry and on  the scaling dimension of $\psi$, $\, \Delta_\psi \equiv (\nu + 1)/2$. 
In the case of a Lifshitz geometry, or $AdS_2\times \mathbf {R}^2$ with $\nu>1$, it has the form of a normal FL. In the case of $\nu<1$, one obtains: 
\begin{equation}
G_{\chi\chi} = 
-\frac{a^{\phantom{a}}}{\vert g \vert^2 c\,\omega^\nu - v_F k_{\perp}} \, . 
\end{equation}
In the last equation,  $a$ is a real parameter whereas $c$ is, in general, complex. 

We must verify that the perturbative corrections do not destabilize the IR fixed point. 
We consider the one-loop corrections to the self-energy in the Supplementary Material. They turn out to be  subleading for small $\omega$, except in the cases of $\nu = 1/2$ and $\nu=1$. We leave the analysis of these cases to future work. 

{\it Generalized Landau-Silin equations.}  We follow the treatment presented in Ref. \cite{Nozieres}, mostly sketching the logic of the derivation and leaving the details for the supplementary material. 
We want to calculate the response of the 
system to an external perturbation, e.g. an electric field which couples to a fermion bilinear. The coupling is obtained by a term in the Lagrangian 
\begin{equation}\label{electric}
\Delta {\cal L} = \int A_\mu ( \chi^\dagger \gamma^\mu \chi + N^2 J^\mu) \,,
\end{equation} 
where $J^\mu$ is a $U(1)$ current of the CFT theory. The observable quantities are also fermion bilinears, so the problem reduces to the computation of a four-point function. We use the notation $\Psi^A = (\chi, \psi)$. The coupling to an external electric field can be read from Eq. (\ref{electric}) and is  $e^{AB}_{\mathbf{k}}( \mathbf{q},t) n^{AB}_{\mathbf{k}}(\mathbf{q}, t)$, with $n^{AB}_{\mathbf{k}}(\mathbf{q}, t)= \Psi^{A\dagger}_{\mathbf{k}-\mathbf{q}/2}(t)\Psi^{B}_{\mathbf{k}+\mathbf{q}/2}(t)$ being the matrix-valued number density operato,r and $e^{AB} = U \, \textrm{diag}(1,N^2)$ with $U$ the electric potential. Notice that we extended the $N$ counting to the coupling with the external field. We might also consider more generally a nondiagonal matrix of external perturbations  
$e^{AB}_{\mathbf{k}}( \mathbf{q},t)$ in which case the  $N$-counting should be $e^{11} = O(1), e^{12}, e^{21} = O(N)$, \text{and} $e^{22}=O(N^2)$ in order to be consistent with large $N$ perturbation theory.

In general, the $U(1)$ current  could have other contributions of the form $ {\cal{\bar O O}}$ with ${\cal O}$ some other charged operator in the CFT. To simplify the analysis, we will assume that these operators are less relevant, meaning that the corresponding bulk fields are more massive than $\psi$ and $\phi$. 

\begin{figure}[h!]
\includegraphics[width=0.47\textwidth]{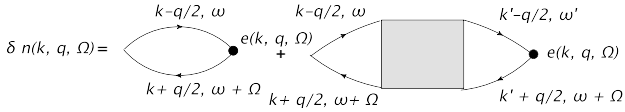}
\caption{Response of the density of quasiparticles in phase
space to an external perturbation.}
\label{integral1}
\end{figure}

The response to an external field is shown diagrammatically in Fig. \ref{integral1}. 
The figure represents the resummation of all the diagrams; the lines are the exact two-point functions, and the shaded box is the exact four-point function. The first term is determined just by the two-point functions, and the second term contains the contribution of the four-point function that can be interpreted as a renormalization of the external field. 
We are interested in the regime when the external field varies slowly in space and time, i.e. $\bq,\Omega \to 0$ (see Fig. \ref{integral1}. In this regime we need to understand the infrared singularities that come both from the two-point and the four-point function $\Gamma$. 

For this purpose, it is convenient to start by considering the sum of the two-particle irreducible diagrams contributing to the four-point function; we call it $\Gamma_0$. In the case of long-range interactions, one can also isolate further the proper part $\tilde \Gamma_0$  given by diagrams that cannot be cut by removing a single bosonic line of interaction; then we have 
\begin{eqnarray*}
\Gamma_0(k,k',q) &=& \tilde \Gamma_0 (k,k',q) + {\cal V}(q) \,, \\
{\cal V}(q) &=& V(q) + \langle \phi(q) \phi(-q) \rangle \,.
\end{eqnarray*}  
where $k = (\bk, \omega), \, k = (\bk', \omega'), \, \text{and} \, q = (\bq, \Omega)$. 
$\cal V$ denotes the full long-range interaction coming from the direct Coulomb interaction of $\chi$ and from the exchange of bulk bosonic fields.  The relation between the full four-point function $\Gamma$ and $\tilde \Gamma_0$ is given by the functional (Bethe-Salpeter) equation represented graphically in Fig. \ref{integral2}.
\begin{widetext}
\begin{center}
\begin{figure}[h!]
\includegraphics[width=0.8\textwidth]{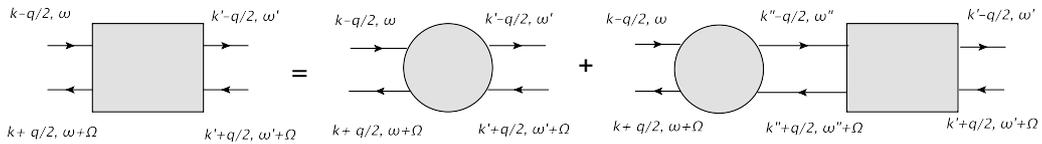}
\caption{Matrix-valued Bethe-Salpeter equation in which boxes represent $\tilde{\Gamma}^{ABCD}$ and circles $\tilde{\Gamma}_0^{ABCD}$}
\label{integral2}
\end{figure} 
\end{center}
\end{widetext}

The basic observation is that the infrared singularity for $\bq,\Omega \to 0$ arises from diagrams in which two $\chi-\chi$ propagators can go on shell while having momenta on the Fermi surface and thus can become singular simultaneously. 
As one can see, the singularity appears in both terms of Fig. (\ref{integral1}). Moreover, this implies that the proper part $\tilde \Gamma_0$ is nonsingular; therefore the Bethe-Salpeter equation relates the complete four-point function to its non-singular part. 
 
We write the product of two two-point functions as  
\begin{eqnarray}
G^{AB}(\mathbf{k}-\mathbf{q}/2, \omega)G^{CD}(\mathbf{k}+\mathbf{q}/2, \omega+ \Omega) = && \nonumber \\ 
a^2 \, Q^{ABCD} (\mathbf{k},\omega, \mathbf{q}, \Omega)+ g^{ABCD}(\mathbf{k}, \omega, \mathbf{q}, \Omega),
\end{eqnarray}
where we isolate $Q^{ABCD}$ which is the singular contribution when $\bq, \Omega \to 0$ (the orders of limits do not commute as in the Fermi liquid case). It depends only on the form of $G(\bk, \omega)$ near the Fermi surface. 
Since there is no such infrared Fermi-surface singularity for the $\psi-\psi$ and $\psi-\chi$ propagators, the singular part has only the $Q^{\chi\chi\chi\chi}$ component. In the Supplementary Material we give details for determining the function $Q$.

Considering the singular contribution in the infrared limit of the equations represented by Figs. \ref{integral1} and \ref{integral2}, we obtain the generalized
Landau-Silin equations in $2+1-d$: 

\begin{eqnarray}\label{landausilin}
\delta n^{AB}_{k}(q) &=&\int \rm{d}^2 k' h^{AB\chi\chi}(\mathbf{k_*}, \mathbf{k'_*})Q^{\chi\chi\chi\chi}(k', q) E_{\mathbf{k'_*}}^{\chi\chi}(q),\nonumber\\
E^{\chi\chi}_{\mathbf{k_*}}(q) &=& \int \rm{d}^2 k' h^{\chi\chi AB}(\mathbf{k_*}, \mathbf{k'_*})e^{AB}_{\mathbf{k'_*}}(q) \\\nonumber &+& \int \rm{d}^2 k' \, [f(\mathbf{k_*}, \mathbf{k'_*})+ \mathcal{V}(q)]Q^{\chi\chi\chi\chi}(k', q)E^{\chi\chi}_{\mathbf{k'_*}}(q),
\end{eqnarray}
%
%
where $k = (\bk, \omega), \, q = (\bq, \Omega)$, 
and $\mathbf{k}_*$ and $\mathbf{k}'_*$ are the points on the Fermi surface nearest to $\mathbf{k}$ and $\mathbf{k}'$, respectively. 
From these equations, we see that $hQ$ can be interpreted as the response function to a renormalized external field $E$. Note that for a generic perturbation, both $\chi$ and $\psi$ bilinears are affected in the low-energy response.

The function on the Fermi surface $f$ gives the generalized Landau parameters and is related to the 4-point function as  
\begin{equation*}
f(\mathbf{k}_*, \mathbf{k}'_*) = a^2 \, \lim_{\substack{ \,\, \bq \to 0\\ \Omega \to 0}} \Gamma^{\chi\chi\chi\chi}(\bk_*, \omega=0, \bk'_*, \omega'=0, \bq, \Omega) \,. 
\end{equation*}
The order of limits ($\bq$ goes to zero first) is important above. In fact, $f$ can be defined in terms of $\tilde \Gamma_0$, as shown in the Supplementary Material. Assuming rotational invariance, $f$ depends only on the angle between $\bk$ and $\bk'$. Using a multipole expansion, we have an infinite number of parameters $F_l$; in practice, one usually assumes that only the first few multipoles are significant.  For microscopic definition of the parameters $h^{AB\chi\chi}$ and $h^{\chi\chi AB}$, we refer the reader to the Supplemental Material.

We can derive all the low-energy phenomenology from Eqs. (\ref{landausilin}). In particular, 
the collective excitations are given by solutions not sourced by an external field, so with $e=0$. 

The main difference from the normal FL case is in the structure of the singular function $Q$. 
In the case of  
$AdS_2 \times {\mathbf R}^2, \, 0 < \nu < 1, \nu \neq 1/2$ there is a genuinely non-Fermi liquid. The function $Q$ in this case is given by 
\begin{widetext}
\begin{eqnarray}\label{singular}
Q &=& \Theta (- k_\perp) \delta(\omega) \frac{1-\nu}{\nu^2} \frac{1}{v_F^2} \left( \frac{v_F}{\lvert g \rvert^2  c} \right)^{1/\nu} k_\perp^{\frac{1}{\nu}-2} 
\frac{q \cos \theta}{q \cos \theta - \nu \Omega (\frac{\lvert g \rvert^2  c}{v_F})^{\frac{1}{\nu}} k_\perp^{1-\frac{1}{\nu}}} \,.
\end{eqnarray}
\end{widetext}
Notice the $\Theta$ function (instead of a delta function on the Fermi surface as in the Landau FL) and the nontrivial dependence on $k_\perp$ both in the numerator and denominator. 
We can then derive the phenomenological consequences of these formulas. In the case of a neutral liquid, ${\cal V}=0$. Then the equation for the zero sound has a solution if the integral operator schematically given by $\int f \, Q $ has an eigenvalue equal to 1. This can happen, for a given $\bq$, only for certain values of $\Omega$; each such value corresponds to a collective mode.  
Assuming for simplicity that $f$ has only a monopole part $F_0$, we find a solution for the zero sound with a dispersion relation $\Omega = v_0 \lvert \bq\rvert$ with velocity $v_0$ determined by the equation 
\begin{eqnarray}\label{zerosound}
\frac{\nu v_F \lvert g \rvert^2  c}{F_0}=  \left( \frac{v_F}{\lvert g \rvert^2  c} k_c \right)^{\frac{1-\nu}{\nu}} - \frac{\nu \lvert g \rvert^2  c v_0}{v_F}\nonumber\\ \textrm{arcsin} \left( \frac{v_F}{\nu \lvert g \rvert^2  c v_0} \left( \frac{v_F}{\lvert g \rvert^2  c} k_c\right)^{\frac{1-\nu}{\nu}} \right) \,
\end{eqnarray} 
for $2+1-d$ non-Fermi liquids. Notice that only filled states below the Fermi surface contribute to the integral; therefore, there is a natural cutoff  $k_c \sim k_F$. 
 The solution is in general complex, so we predict that the zero mode will be present as a peak in the response function but it will be rather damped; moreover 
its dependence on the cutoff shows that it is not a collective excitation only of modes close to the Fermi surface. 
Some universal feature can be obtained in the limit of large $F_0$: the equation shows that $v_0 \sim c^{-3/2\nu} \sqrt{F_0}$, up to real cutoff-dependent prefactors. The ratio of the velocity and attenuation  
is then independent of the cutoff in this limit, and the width of the zero sound, to the extent that it can be defined, is related to the lifetime of the quasiparticles. 

In the charged liquid, with a Coulomb interaction $V(\bq) \sim V_0 / \bq^2$, as usual, the zero sound becomes gapped and there is a plasma frequency. The latter can be obtained by replacing $v_0$ in Eq. (\ref{zerosound}) by $\Omega/ \lvert q \rvert$ and $F_0$ by the Coulomb potential, while expanding in $\lvert q \rvert/ \Omega$. The plasma frequency $\omega_p$ is then found to be:
\begin{equation}
\omega_p^2 = \frac{a^2 V_0}{6 \nu^3 v_F^2} \left( \frac{v_F}{\lvert g \rvert^2 c} \right)^{3/\nu} k_c^{\frac{3}{\nu}-3}
\end{equation}
in $2+1-d$ non-Fermi liquids. We note that this is generically complex with the ratio of the real to imaginary parts again independent of the cutoff. The nonvanishing imaginary part of the plasma frequency denotes that the medium does not act as a sharp filter for the lower frequencies. It would be interesting to explore further the optical properties of our systems by computing the dielectric constant and the refractive index, along the lines of Ref. \cite{Amariti:2010jw}.

In the other cases of $AdS_2 \times {\mathbf R}^2, \nu > 1$ and $AdS_4$/Lifshitz,  the structure of the singular function is similar to that of a Fermi liquid. However, in the charged case, the collective mode  will be radically different if the bulk $\phi-\phi$ exchange gives a force which goes like $1/ q^{2+\alpha}$, with $\alpha > 0$. The plasma frequency will then behave like $q^{-\alpha}$ for small $q$, which means that even high frequency modes will be damped unlike in charged Fermi liquid. Even though we cannot trust the analysis for large $\Omega$, this indicates a different behavior than the Fermi liquid.  

{\it Discussion} We have shown the differences from the predictions of Landau theory that arise in the semiholographic fermion models at the level of collective excitations. 
It would be interesting to see whether the large separation of the zero sound pole from the continuum of two-particle excitations, that is a feature of Landau Fermi liquid theory for large values of $F_0$, is generically true in the class of models we consider. More generally the response of the system to external perturbations needs to be understood in more detail. 

There are a number of directions in which our work can be extended using similar techniques. We plan to compute the thermodynamics and transport properties of the system, as well as studying the system at finite temperature, and the onset of the superconducting instability generalizing BCS theory (see e.g. Ref. \cite{phillips}). 

We should also study the interplay between the collective excitations and quantum kinetics of semiholographic fermions with the hydrodynamic and relaxational modes of the CFT degrees of freedom. This can be pursued using methods developed for computing holographic nonequilibrium spectral and statistical functions \cite{qk}.

For another line of development, it is possible to add Chern-Simons-like term for gauge fields in the bulk. These naturally reproduce global anomalies for the holographically dual CFT. This gives us a concrete method of introducing chiral anomalies in this class of models. 

Finally, it should be possible to apply our approach to the more general class of hyperscaling violating bulk geometries \cite{hyper}. These infrared geometries can be characterized by the dynamical critical exponent $z$ and hyperscaling violating exponent $\theta$.

{\it Acknowledgements.} 
The research of AM is supported by the LABEX P2IO, the ANR Contract No. 05-BLAN-NT09-573739, the ERC Advanced Grant No. 226371 and the ITN Programme No. PITN-GA-2009-237920.
We thank T. Faulkner, A. Garcia-Garcia,  J. Hartong, S. Hartnoll, S. S. Lee, R. Leigh, K. Schalm and J. Zaanen for useful discussions.  



\newpage
\newpage

\section*{SUPPLEMENTAL MATERIAL}

\subsection{Diagonalization of the kinetic term} 

If we consider only the quadratic terms, calling $G_0^{-1}$ the kinetic operator of $\chi$ and $A^{-1}$ that of $\psi$, 
 the fermionic lagrangian in the main text is diagonalised by the change of basis
\begin{eqnarray} 
\eta_1 &=& \psi + \frac{g A}{N} \chi \nonumber \\
\eta_2 &=& \chi - \frac{g^* A}{N} \psi \nonumber
\end{eqnarray}  
with propagators respectively $A/N^2$ and $(G_0^{-1} - \lvert g \rvert^2  A)^{-1}$.
Up to higher order corrections the relation is inverted by 
\begin{eqnarray} 
\chi &=& \eta_2 + \frac{g^* A}{N} \eta_1 \nonumber \\
\psi &=& \eta_1 - \frac{g A}{N} \eta_2 \nonumber
\end{eqnarray}  
Using these relations we immediately find the resummed propagators 
\begin{eqnarray} 
\langle \bar\chi \chi \rangle &=& \frac{1}{G_0^{-1}- \lvert g \rvert^2  A} \nonumber \\
\langle \bar\psi \psi \rangle &=& \frac{1}{N^2} \frac{G_0^{-1} A}{G_0^{-1}- \lvert g \rvert^2  A} 
\end{eqnarray} 
\\
which gives the result of the table in the main text. 
\vskip 8pt

\subsection{Self-energy corrections near the Fermi surface}
In the large N limit, the diagrams contributing to the self-energy are such that the connected pieces in the bulk involves only bulk propagators, because bulk loops and bulk vertices are suppressed. Still one can integrate out the bulk degrees of freedom only perturbatively in our generic model.

\begin{figure}[h!]
\includegraphics[width=0.5\textwidth]{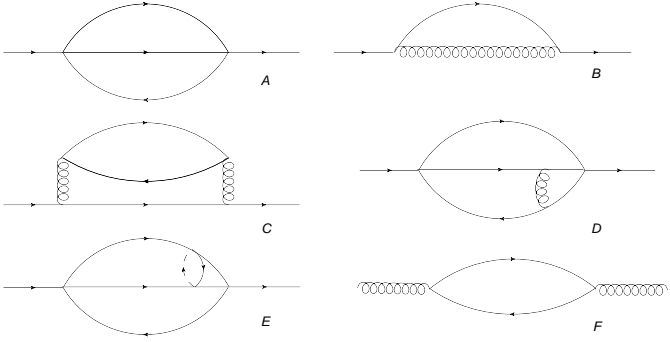}
\caption{Self-energy corrections. The solid lines denote the $\chi-\chi$ propagator, the wavy lines denote long range interaction mediated by bosonic bulk lines/ Coulomb force and the dashed line is the $\psi-\psi$ propagator.}
\label{selfenergy}
\end{figure}

The first few self-energy corrections are as in Fig. \ref{selfenergy}. 
In a neutral fluid where long-range interactions are absent, only diagrams A and E contribute.

Let us evaluate diagram A first. It will be convenient to define the fermion polarizability $\Pi$ first as below:

\begin{equation*}
\Pi (\mathbf{q}, \omega) = \int \frac{\mathrm{d}^2 k}{(2 \pi)^2} \int \frac{\mathrm{d}\epsilon}{2\pi}\, G(\mathbf{k}+ \mathbf{q}, \epsilon + \omega)G(\mathbf{k}, \epsilon).
\end{equation*}

The self-energy contribution of diagram A is then given by:

\begin{equation}\label{self1}
\Sigma(\mathbf{k}, \epsilon) = \lambda^2 \int\frac{\mathrm{d}^2 q}{(2 \pi)^2} \int \frac{\mathrm{d}\omega}{2\pi}\, \Pi(\mathbf{q}, \omega)G(\mathbf{k}+ \mathbf{q}, \omega + \epsilon),
\end{equation}
where for simplicity we have assumed that $\lambda$ is independent of $\mathbf{k}$ and $\mathbf{q}$.

It is useful to consider a local patch around a point on the Fermi surface where we can establish Cartesian coordinates. Let $x$ be the coordinate along the Fermi surface and $y$ be the coordinate perpendicular to it. 
Near the chosen point $\lvert k_\perp\rvert \sim k_y$, and we have 
\begin{equation*}
G^{-1}(\mathbf{k}, \omega= 0) = v_F k_x + \frac{\kappa}{2}k_y^2 \,. 
\end{equation*}

In order to understand the infrared behavior of the self-energy we can consider the Fermi surface to be flat, or in other words push $k_F$ to infinity \cite{SachdevBook} and thus simplify the integrals considerably. Furthermore, we can also integrate over loop momenta and then over the loop frequency in any order as it does not affect the infrared behavior of the integrated result  \cite{SachdevBook}.

Let us first evaluate $\Pi(\mathbf{q}, \omega)$. After appropriate Euclidean continuation of the frequencies, it is given explicitly by:
\begin{widetext}
\begin{eqnarray*}
\Pi (\mathbf{q}, \omega) = \int \frac{\mathrm{d} k_y}{2 \pi} \int \frac{\mathrm{d}\epsilon}{2\pi} \int \frac{\mathrm{d} k_x}{2 \pi} \frac{1}{\lvert g\rvert^2 c (\epsilon + \omega)^\nu + i v_F (k_x + q_x) + i \frac{\kappa}{2}(k_y + q_y)^2} \frac{1}{\lvert g\rvert^2 c \epsilon^\nu + i v_F k_x  + i \frac{\kappa}{2}k_y^2}.
\end{eqnarray*}
\end{widetext}
Note the specific order of integration  chosen in order to simplify the integration. The $k_x$ integral can be performed using residues, with the result 
\begin{equation*}
\Pi  =  \frac{1}{h v_F} \int \frac{dk_y}{2\pi} \frac{d\epsilon}{2\pi} \frac{\text{sgn}(\text{Re}(\epsilon+\omega)^\nu)-\text{sgn}(\text{Re} (\epsilon)^\nu)}{(\epsilon+\omega)^\nu - \epsilon^\nu + \frac{i \kappa}{h} (k_y q_y + \frac{q_y^2}{2}) + \frac{i v_F}{h}q_x}
\end{equation*}
where $h = \left| g \right|^2 c$, and we have used the fact that for a neutral liquid, $c$ is real. 

Then considering that 
\begin{eqnarray*}
\text{sgn} (\text{Re}\,(x^\nu)) = \begin{cases}
\text{sgn} (x) \,\,\, \text{if} \,\,\, 1/2 < \nu < 1 \\
1 \,\, \, \text{if} \,\,\, 0 < \nu < 1/2
\end{cases}
\end{eqnarray*}
one readily sees that the polarization integral vanishes for $0 < \nu < 1/2$. 
For the other case $1/2 < \nu < 1$, we can approximate in the denominator $(\epsilon+ \omega)^\nu - \epsilon^\nu \approx \omega^\nu$, then the $\epsilon$ integral can be immediately performed, since  
$$\int \mathrm{d}\epsilon (\text{sgn}(\omega + \epsilon) - \text{sgn}(\epsilon)) = 2 \omega \,,$$ 
and we find 
\begin{equation*}
\Pi(\mathbf{q}, \omega) \approx 
- \frac{\omega}{\pi v_F}\int \frac{\mathrm{d}k_y}{2\pi} \, \frac{1}{h  \omega^\nu + i v_F q_x + i \frac{\kappa}{2}q_y^2 + i \kappa q_y k_y }
\end{equation*}
The final $k_y$ integral can again be done by residues: 
\begin{equation*}
\Pi(\mathbf{q}, \omega) \approx 
\frac{1}{\pi v_F \kappa} \frac{\omega}{q_y} \text{sgn} \frac{\text{Re}(\omega^\nu)}{q_y} = \frac{1}{\pi v_F \kappa} \frac{\lvert \omega \rvert}{\lvert q_y \rvert} \,. 
\end{equation*}
In summary, the polarization vanishes for $0 < \nu < 1/2$, and it has the same form as in the Fermi liquid for 
$1/2 < \nu < 1$. 

Finally we obtain the self energy correction using Eq. (\ref{self1}). Clearly it follows from above that $\Sigma (\mathbf{k}, \epsilon)$ vanishes for small $\mathbf{k}$ and  $\omega$ when $0< \nu < 1/2$. When $1/2 < \nu < 1$, we get
\begin{eqnarray*}
\Sigma (\mathbf{k}, \epsilon) = - \lambda^2 \int \frac{\mathrm{d} q_y}{2 \pi} \int \frac{\mathrm{d}\omega}{2\pi} \int \frac{\mathrm{d} q_x}{2 \pi}\\
\frac{\lvert\omega \rvert}{\pi v_F\kappa \lvert q_y \rvert} \frac{1}{h  (\epsilon + \omega)^\nu +i v_F(k_x + q_x) + i \frac{\kappa}{2}(k_y+ q_y)^2}
\end{eqnarray*}

We can readily do the $q_x$ integral by taking the pole contribution. We obtain
\begin{eqnarray*}
\Sigma (\mathbf{k}, \epsilon) &\approx&  \lambda^2  \int\frac{\mathrm{d} q_y}{2 \pi} \int \frac{\mathrm{d}\omega}{2\pi}  
\frac{\lvert\omega \rvert}{\pi v_F\kappa \lvert q_y \rvert} \text{sgn}(\epsilon + \omega)\\
&\approx&  \frac{\lambda^2 \epsilon^2}{2\pi^2 v_F^2 \kappa }\text{sgn}(\epsilon)\int\frac{\mathrm{d} q_y}{2 \pi}\frac{1}{\lvert q_y \rvert}\\
&\approx& \frac{\lambda^2 \epsilon^2}{2\pi^2 v_F^2 \kappa }\text{sgn}(\epsilon)\log \Bigg(\frac{k_c}{\lambda\lvert \omega \rvert}\Bigg).
\end{eqnarray*}
In the last step we have used the fact that the logarithmic integral is automatically regulated in the manner shown if the damping is introduced in a self-consistent manner  \cite{SachdevBook}. Remarkably the self-energy corrections are very much like a Fermi liquid and subleading at the Fermi surface. These  results are compatible with the intuition one can develop from renormalization group arguments. Thus one can argue that even higher order perturbative corrections will be subleading at the Fermi surface.

A problem one encounters wen trying to couple a Fermi liquid to a critical boson \cite{SSLee} is that the fermion polarization will renormalize the boson propagator (diagram F in Fig. \ref{selfenergy}); since it is a singular contribution, it becomes more important than the bare propagator in the IR; in turn, the one-loop corrected boson propagator affects the fermion one by diagram B, and the IR fixed point is destabilized. In our case, the situation is actually better. Even though diagram F is not suppressed in the $N$-counting, the effect is a correction $\lvert \omega \rvert$ to a bare term of order $\omega^\nu$, so it does not destabilize the fixed point for $0 < \nu < 1$, at least by a naive application of the RG argument. However it would be important to confirm this intuition by careful computations.

%

\subsection{Scaling RG arguments} 

In this section we consider the RG scaling of the terms in our model in order to determine what are the relevant and marginal perturbations.  We follow the analysis of Ref. \cite{Polchinski:1992ed}, where it has been pointed out that the RG around a Fermi surface is special as the momenta are rescaled towards the surface and not towards the origin; in a Wilsonian approach, in order to change the cutoff from $\Lambda$ to $s \Lambda$  one  integrates out modes with momenta $ s \Lambda < \lvert k-k_F \rvert < \Lambda$. 
If we write the fermion momentum as ${\bf p} = {\bf k} + \bl$, with ${\bf k}$ on the Fermi surface and ${\bf l}$ orthogonal to it, then the scaling is $[{\bf k}] = 0, [{\bf l}] = 1,$ where $[A] = n$ means that it scales as $A \to s^n A$. 
 In order to have all the kinetic terms to be scale-invariant one then finds 
$$[ t ] = -1 \,, \quad [\chi] = - \frac{1}{2} \,. \quad $$
On the other hand, for the bosonic and fermionic operators of the bulk CFT, if we consider the case of $AdS_2$ infrared geometry, the momentum does not scale since the correlator is only a function of $\omega$; then one finds 
$$[\phi] = [\psi] = \frac{\nu-1}{2} \,.$$
Let us now consider an interaction term $(\chi^\dagger \chi)^n \phi$ or  $(\chi^\dagger \chi)^n \psi$ (in the latter case we take $n$ half-integer, with a slight abuse of notation).  Written in momentum space the interaction contains a delta function for momentum conservation, 
$$\int dt dk_i d p \chi_{k_1} \ldots \chi_{k_n} \phi_p \delta^d(\sum_i k_i + p)\,. $$
For a generic kinematic configuration, where the sum of the fermionic momenta does not lie on the Fermi surface, the delta function does not scale. Then one finds that the dimension of the interaction is $n-1 + \frac{\nu-1}{2}$, so this is relevant or marginal if 
$$n \leq \frac{3-\nu}{2} \, .$$
Considering that $0 < \nu <1$ one sees that the only interactions that are generically marginal or relevant are for $n=1$ and $n=3/2$. These imply we need to consider the interactions $(\chi^\dagger \chi) \phi$, $(\chi^\dagger \chi)\chi^\dagger \psi$ and it's complex conjugate only.

\subsection{Derivation of the singular part \\ of the product of propagators}

Let us consider first the case of a normal Fermi liquid. The Feynman's Green function is given by 
\begin{equation}
G(k,\omega) = \frac{1}{\omega - v_F k_\perp + i \delta \sigma(k_\perp)}
\end{equation} 
where $\sigma$ is the sign function and $\delta = 0^+$. This propagator is characterized by  the property of having poles in the lower half-plane (LHP) of complex frequencies for $k_\perp > 0$ and in the upper half-plane (UHP) for $k_\perp < 0$; 
moreover, it is equal to $G_R$ for $\omega>0$ and to $G_A$ for $\omega<0$. \\
When we consider the product of two Green's functions, $G(k) G(k+q)$, and we perform the integral over $\omega$ on a contour that includes the real axis and the circle going to infinity in the UHP, there are four possibilities: 
\begin{itemize}
\item $k_\perp >0, (k+q)_\perp > 0$ : no poles in the UHP , zero contribution
\item $k_\perp < 0, (k+q)_\perp <0$ : two poles in the UHP with opposite residues , zero contribution
\item $k_\perp < 0, (k+q)_\perp >0$ or $k_\perp > 0, (k+q)_\perp <0$ : one pole in the UHP 
\end{itemize}
so the integration gets contribution only in a range $k_\perp \lesssim |q|$. As $q \to 0$ the product can be approximated by $ A \delta(\omega) \delta (k_\perp)$. The result is 
\begin{equation*}
Q = \delta(v_F k_\perp)\delta(\omega)a^2 \frac{v_F q \cos\theta}{\Omega - v_F q \cos\theta} \,.
\end{equation*}

Let us now consider the case of the non-Fermi liquid. The retarded Green's function has the form
\begin{equation} 
G_R = \frac{a}{\omega + h \, e^{i \gamma} \, \omega^\nu - v_F k_\perp}
\end{equation} 
where $a, h >0$ and we consider the charged case, so we explictly included the phase factor in the self-energy, so in this section $\lvert g \rvert ^2 c= h e^{i \gamma}$. One can show that $0 < \gamma < \pi (1 - \nu)$.  The behaviour of the poles of this function is illustrated in Fig. 2 in Ref. \cite{Faulkner:2009wj}. 
Because of the term with non-integer power the function has a branch cut that can be chosen to run along the negative imaginary axis. 
First we must find the Feynman propagator to use in the diagrams. By analogy with the free case, we find that the right prescription is obtained with the following form: 
\begin{equation} 
G(k,\omega) = \frac{a}{\omega + h \, e^{- i \gamma \sigma(k_\perp)} \, \omega^\nu - v_F k_\perp}
\end{equation}
To perform the integral, we map $\omega \to \zeta = \omega^\nu$. This maps the UHP into a wedge of angle $\pi \nu$. The integral is along the edges of the wedge and the 
contour at infinity that does not cross the branch cut. 
Again we have four possibilities for the location of the poles, but now in case 2, when both poles are inside the contour, there is no cancellation. In cases 3 and 4, where the integration 
is for momenta close to the Fermi surface, one finds that the residue of the pole vanishes like $k_\perp^{(1/\nu - 1)}$, so there is no contribution. 
The result is then 
\begin{widetext}
\begin{equation*}
\theta(-k_\perp) \theta(- (k + q)_\perp) \frac{a^2}{\lvert g \rvert^4 c^2} \int d\omega \, \frac{1}{\omega^\nu + \frac{v_F}{\lvert g \rvert^2 c} |k_\perp| } \frac{1}{(\omega+ \Omega)^\nu + \frac{v_F}{\lvert g \rvert^2 c} |(k+q)_\perp|} 
\end{equation*}
\end{widetext}
where $\lvert g \rvert^2 c = h e^{i \gamma}$. Using 
$$ \int \frac{d \omega}{2 \pi} \frac{f(\omega)}{\omega^\nu -a} = \frac{i}{\nu} a^{\frac{1-\nu}{\nu}} f(a^{1/\nu}) $$
we find 
\begin{equation*} 
\theta(- k_\perp) \frac{i a^2}{\lvert g \rvert^4 c^2}\frac{1-\nu}{\nu^2} (\hat k_\perp)^{\frac{1-2\nu}{\nu}} \, \frac{\hat q \, \cos \theta }{\hat q \, \cos \theta - \nu \Omega (\hat k_\perp)^{1-\frac{1}{\nu}}}
\end{equation*} 
where $(\hat k, \hat q) = \frac{v_F}{\lvert g \rvert^2 c}( k,q)$. \\
Integrating over $\theta$ and $k_\perp$ (this integral needs a UV cutoff $k_c$) we find 
\begin{equation*} 
\frac{i a^2}{v_F \nu \lvert g \rvert^2 c} \big( \hat k_c^{\frac{1}{\nu}-1} - \frac{\nu \Omega}{\hat q} \textrm{arcsin}\left( \frac{\hat q}{\nu \Omega} \hat k_c^{\frac{1}{\nu}-1} \right) \big) 
\end{equation*}
From the last two results the equations (6) and (7) of the main text follow. 

\subsection{Derivation of Landau-Silin equations}

In this section we review the derivation of the Landau-Silin equations generalized to our system. 
Let us begin with the case of uncharged Fermi liquid where long range forces are absent. The Bethe-Salpeter equation written explicitly is as below:
\begin{widetext}
\begin{eqnarray}\label{BSeq}
\Gamma^{ABCD}(k; k'; q) &=& \Gamma_0^{ABCD}(k; k'; q) \, \\
&&+\int dk'' \Gamma_0^{ABEF}(k, k'', q)
G^{EG}(\mathbf{k}''-\mathbf{q}/2, \omega'') G^{FH}(\mathbf{k}''+\mathbf{q}/2, \omega''+ \Omega) \Gamma^{GHCD}(k'', k', q) \, , \nonumber
\end{eqnarray} 
\end{widetext} 
with $\Gamma$ is the full matrix-valued four-point function and $\Gamma_0$ is its fermionic two-particle irreducible part.

It is easy to see that the transfer of energy-momentum $q$ from one fermionic pair to another is merely a spectator in the integral equation. We may regard $\Gamma$ and $\Gamma_0$ as matrices in flavor space (with two flavor indices forming the row entries and two flavor indices forming the column entries) and $k$ and $k'$ space. Similarly the two-particle bridge $GG$ can also be regarded as a diagonal matrix in $k$ and $k'$ space, and a non-diagonal matrix in flavor space. We recall, as used in the main text, that  $GG$ can be split as $GG = a^2 Q + g$ with $Q$ being singular and $g$ being non-singular as $q$ vanishes. We recall $a$ is just a number related to the residue of the pole of the $\chi-\chi$ propagator at the Fermi-surface in $\omega^\nu$. The matrix multiplication can then be obviously defined.

In the matrix language, the spatial Fourier transform of the Wigner function representing phase space density operator of fermions given by $\hat{n}^{AB}_\mathbf{k}(\mathbf{q}, t) = \hat{c}^{A\dagger}_{\mathbf{k}+ \mathbf{q}/2}(t) \hat{c}^B_{\mathbf{k}-\mathbf{q}/2}(t)$, can be thought of as a vector. As self-energy corrections are irrelevant at the Fermi surface, we can write any relevant perturbation as $\hat{n}e$ as observed in the main text and similarly any observable as $p^{\dagger}\hat{n}$, with $e_{\mathbf{k}}^{AB}(\mathbf{q}, t)$ and $p_{\mathbf{k}}^{AB}(\mathbf{q}, t)$ representing appropriate vectors. The inner product $\hat{n}e$ and $p^{\dagger}\hat{n}$ involves integration over $\mathbf{k}$, and summation over $A$ and $B$; the $\mathbf{q}$ and $t$ dependence then captures the space-time variation of the perturbation and response respectively. Note more generally if the perturbation is of the form
\begin{equation*}
 \int \mathrm{d} t \int \mathrm{d} t' \, e^{AB}_{\mathbf{k}}(\mathbf{q}, t, t')\hat{c}^{A\dagger}_{\mathbf{k}+ \mathbf{q}/2}(t) \hat{c}^B_{\mathbf{k}-\mathbf{q}/2}(t'),
\end{equation*}
then it is useful to define 
\begin{widetext}
\begin{equation*}
\hat{n}_{\mathbf{k}, \omega}^{AB}(\mathbf{q}, \Omega) = \int \mathrm{d} T \int \mathrm{d} t_r \,\, e^{-i\Omega t_r}e^{-i\omega T}\, \, \hat{c}^{A\dagger}_{\mathbf{k}+ \mathbf{q}/2}(T + t_r/ 2) \hat{c}^B_{\mathbf{k}-\mathbf{q}/2}( T - t_r /2).
\end{equation*}
\end{widetext}
Combining $(\mathbf{k}, \omega)$ into the three-vector $k$ and $(\mathbf{q}, \Omega)$ into three-vector $q$ we can thus define the vector $n_k( q)$. Similarly we can define $e_k(q)$ using
\begin{widetext}
\begin{equation*}
e_{\mathbf{k}, \omega}^{AB}(\mathbf{q}, \Omega) = \int \mathrm{d} T \int \mathrm{d} t_r \,\, e^{-i\Omega t_r}e^{-i\omega T}\, \, e^{AB}_{\mathbf{k}}(\mathbf{q}, T + t_r/ 2, T- t_r/ 2).
\end{equation*}
\end{widetext}
such that the general external perturbation again takes the abridged form $\hat{n}e$, the inner product involving an integration over $k$ and summation over $A$ and $B$. Similarly one can consider a more general observable of the form $p^\dagger \hat{n}$ such that $p$ is a function of both $k$ and $q$.

The diagrammatic equations giving the response of an arbitrary observable to an arbitrary perturbation, illustrated in Fig. 1 in the main text,  can be written in the matrix form as below:
\begin{equation} \begin{split}\label{matrixform}
R =& p^\dagger (a^2 Q+g)\Big[1 + \Gamma(a^2 Q+g)\Big]e, \\
\Gamma =& \Gamma_0 + \Gamma_0(a^2 Q+g) \Gamma ,
\end{split} \end{equation}
where $R$ denotes the response under study.

We note that in all cases 
\begin{equation}\label{Qgt0}
\lim_{\Omega\rightarrow 0, \mathbf{q}\rightarrow 0}Q = 0,
\end{equation}
while taking the limits in the opposite order yields a non-vanishing result. This plays a crucial role in generalizing Landau-Silin theory. Let us denote the non-singular part of the four-point function as:
\begin{equation}\label{Gamma1}
\Gamma_1 = \lim_{\Omega\rightarrow 0, \mathbf{q}\rightarrow 0} \Gamma = \Gamma_0 + \Gamma_0 g \Gamma_1,
\end{equation}
where the second equality follows by using Eq. (\ref{Qgt0}) in the second equation in (\ref{matrixform}). Assuming inverses of matrices exist we can show after some manipulations that
\begin{equation}
\Gamma = (1- \Gamma_0 g)^{-1}\Big[ \Gamma_0 + a^2 \Gamma_0 Q \Gamma \Big] = \Gamma_1 + a^2 \Gamma_1 Q \Gamma. 
\end{equation}
Similar manipulations give us
\begin{widetext}
\begin{eqnarray}\label{matrixform2}
R &=& p^{\dagger}g(1+ \Gamma_1 g)e + a^2 p^\dagger (1+ g\Gamma_1)(1- a^2 Q\Gamma_1)^{-1} Q(1+ \Gamma_1 g)e,\\\nonumber
&=& p^{\dagger}g(1+ \Gamma_1 g)e +a^2 p^\dagger (1+ g\Gamma_1)Q(1- a^2 \Gamma_1 Q)^{-1} (1+ \Gamma_1 g)e,
\end{eqnarray}
\end{widetext}
where we have used
\begin{equation*}
(1- Q\Gamma_1)^{-1} Q = Q(1- \Gamma_1 Q)^{-1}.
\end{equation*}
In the limit $q\rightarrow 0$, we can ignore the first term on the right hand side of Eq. (\ref{matrixform2}). Therefore in this limit,
\begin{equation}
R = a^2 p^\dagger (1+ g\Gamma_1)Q(1- a^2 \Gamma_1 Q)^{-1} (1+ \Gamma_1 g)e.
\end{equation}

We will now derive some useful identities. Consider the case of a perturbation $e$ which is constant, \textit{i.e.} independent of $q$. In such a case, it follows that
\begin{widetext}
\begin{eqnarray}
\delta G(k) &=& \lim_{\Omega\rightarrow 0, \mathbf{q}\rightarrow 0}\Big[G(k) \delta e(q) G(k+  q) - i G(k) G(k+ q)\int \frac{\mathrm{d}^3 k'}{(2\pi)^3}\Gamma(k, k'; q) G(k') \delta e(q) G(k' + q)\Big], \\\nonumber
&=& g(k, q=0)\Big[1+ \int\frac{\mathrm{d}^3 k'}{(2\pi)^3} \Gamma_1(k, k')g(k', q= 0)\Big]\delta e (q=0)
\end{eqnarray}
\end{widetext}
Therefore,
\begin{equation}\label{id1}
\frac{\delta G^{-1}(k)}{\delta e_{k'}} = \Big[1 + \Gamma_1g\Big] (k, k', q= 0)
\end{equation}
It is not hard to see that on the Fermi surface
\begin{equation}
\frac{\delta G^{\chi\chi-1}(\mathbf{k}= \mathbf{k_*}, \omega = 0 )}{\delta e_{\mathbf{k'}_*}^{\chi\chi}} = a^{-1} \delta(\mathbf{k_*} - \mathbf{k'_*})
\end{equation}
Combining the above identity with Eq. (\ref{id1}) we find that
\begin{equation}\begin{split}\label{Wid}
\Big[1 + \Gamma_1g\Big] ^{\chi\chi\chi\chi}(\mathbf{k}= \mathbf{k}_*, & \mathbf{k'}= \mathbf{k'}_*,\omega = 0; q= 0) \\ & = a^{-1}\delta(\mathbf{k_*} - \mathbf{k'_*}).
\end{split}\end{equation}
Ignoring non-singular terms in Eq. (\ref{matrixform2}) it follows that
\begin{widetext}
\begin{eqnarray}
\delta n^{AB}_{k}(q) &=&\int \rm{d}^2 k' h^{AB\chi\chi}(\mathbf{k_*}, \mathbf{k'_*})Q^{\chi\chi\chi\chi}(k', q) E_{\mathbf{k'_*}}^{\chi\chi}(q)\nonumber\\
E^{\chi\chi}_{\mathbf{k_*}}(q) &=& \int \rm{d}^2 k' h^{\chi\chi AB}(\mathbf{k_*}, \mathbf{k'_*})e^{AB}_{\mathbf{k'_*}}(q) + \int \mathrm{d}^2 k' \, f(\mathbf{k_*}, \mathbf{k'_*})Q^{\chi\chi\chi\chi}(k', q)E^{\chi\chi}_{\mathbf{k'_*}}(q),
\end{eqnarray}
where
\begin{equation} 
f(\mathbf{k_*}, \mathbf{k'_*}) = a^2 \Gamma_1^{\chi\chi\chi\chi}(\mathbf{k} = \mathbf{k_*}, \omega = 0, \mathbf{k'} =\mathbf{k'_*}, \omega' = 0),
\end{equation}
and
\begin{eqnarray}
h^{AB\chi\chi}(\mathbf{k_*}, \mathbf{k'_*}) &=& a \Bigg(1+ g\Gamma_1 \Bigg)^{AB\chi\chi}(\mathbf{k} = \mathbf{k_*},\omega = 0, \mathbf{k'} = \mathbf{k'_*}, \omega' = 0; \mathbf{q}= 0, \Omega = 0), \nonumber\\
h^{\chi\chi AB}(\mathbf{k_*},\mathbf{k'_*}) &=& a \Bigg(1+ \Gamma_1 g\Bigg)^{AB\chi\chi}(\mathbf{k} = \mathbf{k_*}, \omega = 0, \mathbf{k'} = \mathbf{k'_*},\omega' = 0; \mathbf{q}= 0, \Omega = 0).
\end{eqnarray}
\end{widetext}
Note it follows from Eq. (\ref{Wid}) that $h^{\chi\chi\chi\chi} = 1$. The integral equation for $f(\mathbf{k_*}, \mathbf{k'_*}) $ follows from Eq. (\ref{Gamma1}). Thus we recover the Landau equations for the neutral non-Fermi liquid. In summary the low energy response to an arbitrary perturbation is characterized by the parameters of the two-point function on the Fermi surface namely $a$, $g$, $c$ and the scaling exponent $\nu$; the generalized Landau parameters $f(\mathbf{k_*}, \mathbf{k'_*})$ obtained from the four-point function; and the parameters obtained from both the non-singular parts of the four-point and the two-point function which are $h^{\chi\chi\chi\psi}(\mathbf{k_*},\mathbf{k'_*})$, $h^{\chi\chi\psi\chi}(\mathbf{k_*},\mathbf{k'_*})$, $h^{\chi\chi\psi\psi}(\mathbf{k_*},\mathbf{k'_*})$, $h^{\chi\psi\chi\chi}(\mathbf{k_*},\mathbf{k'_*})$, $h^{\psi\chi\chi\chi}(\mathbf{k_*},\mathbf{k'_*})$ and $h^{\psi\psi\chi\chi}(\mathbf{k_*},\mathbf{k'_*})$.

We can similarly derive the case of the charged Fermi liquid by following the methodology of Silin 
(see e.g. Ref. \cite{Nozieres}).

\end{document}